\documentclass[preprint,nofootinbib,superscriptaddress,showpacs,amsmath]{revtex4}

\usepackage{graphicx}
\usepackage{dcolumn}
\usepackage{bm}

\usepackage{times}

\usepackage{graphicx}
\usepackage{dcolumn}
\usepackage{bm}

\begin{document}

\title{
Radiative corrections to the three-body region of the Dalitz plot of baryon semileptonic decays with angular correlation between polarized emitted baryons and charged leptons: The initial-baryon rest frame case
}

\author{
C.\ Ju\'arez-Le\'on\
}
\affiliation{
Escuela Superior de F\'{\i}sica y Matem\'aticas del IPN, Apartado Postal 75-702, M\'exico, D.F.\ 07738, Mexico
}

\author{
A.\ Mart{\'\i}nez
}

\affiliation{
Escuela Superior de F\'{\i}sica y Matem\'aticas del IPN, Apartado Postal 75-702, M\'exico, D.F.\ 07738, Mexico
}

\author{
M.\ Neri
}
\affiliation{
Escuela Superior de F\'{\i}sica y Matem\'aticas del IPN, Apartado Postal 75-702, M\'exico, D.F.\ 07738, Mexico
}

\author{
J.\ J.\ Torres
}

\affiliation{
Escuela Superior de C\'omputo del IPN, Apartado Postal 75-702, M\'exico, D.F. 07738, Mexico
}

\author{
Rub\'en Flores-Mendieta
}

\affiliation{
Instituto de F{\'\i}sica, Universidad Aut\'onoma de San Luis Potos{\'\i}, \'Alvaro Obreg\'on 64, Zona Centro, San Luis Potos{\'\i}, S.L.P.\ 78000, Mexico
}

\author{A.\ Garc{\'\i}a
}

\affiliation{
Departamento de F{\'\i}sica, Centro de Investigaci\'on y de Estudios Avanzados del IPN, Apartado Postal 14-740, M\'exico, D.F.\ 07000, Mexico
}

\date{\today}

\begin{abstract}
We complement the results for the radiative corrections to the ${\hat {\mathbf s}_2} \cdot {\hat {\mathbf l}}$ angular correlation of baryon semileptonic decays of Ref.~\cite{neri08} with the final results in the rest frame of the decaying baryon.
\end{abstract}


\maketitle

In a recent paper \cite{neri08} we obtained the radiative corrections (RC) to the Dalitz plot of the semileptonic decay of a spin-1/2 baryon $A$,
\begin{equation}
A^{\substack{-\\0}}(p_1) \to B^{\substack{0\\+}}(p_2) + \ell^-(l) + \overline{\nu}_\ell(p_\nu),
\end{equation}
when the angular correlation ${\hat {\mathbf s}_2} \cdot {\hat {\mathbf l}}$ between the spin ${\hat {\mathbf s}_2}$ of the emitted baryon $B$ and the direction ${\hat {\mathbf l}}$ of the emitted charged lepton $\ell$ is observed. As is customary, the results were presented in the rest frame of $B$ where $p_2=(M_2,0,0,0)$. However, due to experimental conditions, it may be more convenient to produce such RC in the rest frame of $A$ where $p_1=(M_1,0,0,0)$. It is not possible to translate directly the final result of Ref.~\cite{neri08} into the final result of the latter RC. The calculation must be retaken starting at earlier stages. In this paper we shall complement the analysis of Ref.~\cite{neri08} and present the final result for the RC to the ${\hat {\mathbf s}_2} \cdot {\hat {\mathbf l}}$ angular correlation in the rest frame of $A$.

We shall follow the same procedure and use the same conventions and notation of Ref.~\cite{neri08}. We shall omit a detailed discussion, which can be found in this reference. Let us just recall that our results for the virtual part will be model-independent, gauge invariant, finite in the ultraviolet, and will contain the infrared divergence. In order to avoid repetition of long expressions, it will be convenient to trace a close parallelism with the analysis of the RC for the angular correlation ${\hat {\mathbf s}_1} \cdot {\hat {\mathbf l}}$ between the spin ${\hat {\mathbf s}_1}$ of $A$ and the
direction ${\hat {\mathbf l}}$ of $\ell$ which were obtained in Ref.~\cite{mar01}.

Without further ado, the differential decay rate with virtual RC in the rest frame of $A$ including the ${\hat {\mathbf s}_2} \cdot {\hat {\mathbf l}}$ correlation and covering the two charge assignments of $A^{\substack{-\\0}}$ is
\begin{equation}
d\Gamma_V = d\Gamma_V^\prime - d\Gamma_V^{(s)}.
\end{equation}

The unpolarized part was already calculated in Ref.~\cite{mar01}. It has the form
\begin{equation}
d\Gamma_V^\prime = d\Omega \left[ A_0^\prime + \frac{\alpha}{\pi} \left(A_1^\prime\phi + \phi^\prime A_1^{\prime \prime} \right) \right]. \label{eq:dVp}
\end{equation}

The full expressions for $A_0^\prime$, $A_1^\prime$, $A_1^{\prime \prime}$, $\phi$, and $\phi^\prime$ can be found in
Eqs.\ (B1), (B2), (B3), (11), and (12) of this reference. Here the phase space factor $d\Omega = (1/2) (G_V^2/2) dE_2dEd\Omega_{\ell}d\phi_2 2M_1/(2\pi)^5$ is one-half the phase space factor of Eq.~(25) of this same reference. The polarization appears in
\begin{equation}
d\Gamma_V^{(s)} = d\Omega \left[ A_0^{(s)} + \frac{\alpha}{\pi} \left( B^\prime\phi + \phi^\prime B^{\prime \prime} \right) \right] {\hat {\mathbf s}_2} \cdot {\hat {\mathbf l}}. \label{eq:dV}
\end{equation}
${\hat {\mathbf s}_2}$ is introduced with the spin projector $\Sigma(s_2) = (1-\gamma_5 {\not \! s_2})/2$ applied to the $u_{B}$ spinor.
The four vector $s_2$ obeys $s_2 \cdot s_2 = -1$ and $s_2 \cdot p_2 = 0$. The trace calculation will lead to products $s_2 \cdot a$ with $a=l,p_\nu, p_1$. These products are specialized to the rest frame of $A$ using the relation \cite{gins71}
\begin{equation}
s_2 \cdot a = {\mathbf s}_2^A \cdot \left[ \frac{{\mathbf p}_2}{E_2}a_0 - {\mathbf a} \right] = {\hat {\mathbf s}_2}^B \cdot \left[ \frac{{\mathbf p}_2}{M_2} \left(a_0 - \frac{{\mathbf p}_2 \cdot {\mathbf a}}{E_2 + M_2} \right) - {\mathbf a}\right], \label{eq:ginstr}
\end{equation}
which corresponds to the Lorentz transformation from the rest frame of $B$ to the rest frame of $A$. In the first equality of (\ref{eq:ginstr}) it is understood that ${\mathbf p}_2$, ${\mathbf s}_2^A$, and the components of the 4-vector $a=(a_0,{\mathbf a})$ are specialized in the rest frame of $A$. In the second equality of (\ref{eq:ginstr}) only ${\hat {\mathbf s}_2}^B$ still remains in the rest frame of $B$. The upper indices $A$ and $B$ on the spin of the emitted baryon emphasize this fact. In addition, the contribution of the correlations ${\hat {\mathbf s}_2}^B \cdot {\hat {\mathbf p}_2}$ and ${\hat {\mathbf s}_2}^B \cdot {\hat {\mathbf p}_\nu}$ to the ${\hat {\mathbf s}_2}^B \cdot {\hat {\mathbf l}}$ correlation over the Dalitz plot are taken into account with the substitution rule \cite{mar01} ${\hat {\mathbf s}_2}^B \cdot {\hat {\mathbf p}} \to ({\hat {\mathbf s}_2}^B \cdot {\hat {\mathbf l}})({\hat {\mathbf p}} \cdot {\hat {\mathbf l}})$, with ${\hat {\mathbf p}} = {\hat {\mathbf p}_2}, {\hat {\mathbf p}_\nu}$, which is valid under integration of the variables, other than $E$ and $E_2$, contained in $d\Omega$. These steps will be extended to the bremsstrahlung part, where ${\hat {\mathbf p}}={\hat {\mathbf k}}$ will also appear.

The results in Eq.~(\ref{eq:dV}) are new. Their explicit expressions are
\begin{equation}
A_0^{(s)} = Q_1^\prime \left[ p_2 y_0 \frac{M_1}{M_2} \right] + Q_2^\prime \left[ \frac{p_2 y_0}{M_2} \left(E - \frac{
p_2ly_0}{E_2 + M_2} \right) - l \right], \label{eq:aprime}
\end{equation}
\begin{equation}
B^\prime = -D_4 E_\nu^0 l + D_3E (p_2 y_0 +l), \label{eq:bprime}
\end{equation}
\begin{equation}
B^{\prime\prime} = -D_3E {\hat {\mathbf l}} \cdot {\mathbf p}_\nu,
\end{equation}
where
\begin{eqnarray}
Q_1^\prime & = & (F_1^2-G_1^2) \frac{EE_2 - {\mathbf p}_2 \cdot {\mathbf l}}{M_1} - (F_1^2+G_1^2-2F_1G_1) \frac{M_2E}{M_1}  \nonumber \\
&  & \mbox{} + (F_1G_2+F_2G_1) \frac{EM_2^2 + (E-E_\nu^0) (EE_2 - {\mathbf p}_2 \cdot {\mathbf l})}{M_1^2} \nonumber \\
&  & \mbox{} - (F_1G_2-F_2G_1) \frac{M_2(EE_2 - {\mathbf p}_2 \cdot {\mathbf l} + m^2)}{M_1^2} \nonumber \\
&  & \mbox{} + (F_1G_3+F_3G_1) \frac{m^2}{M_1^3} (EE_2 - {\mathbf p}_2 \cdot {\mathbf l} + M_2^2) \nonumber \\
&  & \mbox{} - (F_1G_3-F_3G_1) \frac{M_2 m^2}{M_1^2} + (F_1F_2+G_1G_2) \frac{Ep_2^2 - E_2 {\mathbf p}_2 \cdot {\mathbf l}}{M_1^2} \nonumber \\
&  & \mbox{} - F_2 G_2 \frac{M_2}{M_1^2} (-l^2 - {\mathbf p}_2 \cdot {\mathbf l} + EE_\nu^0) - (F_2G_3+F_3G_2) \frac{M_2 m^{2}E_\nu^0}{M_1^3} \nonumber \\
&  & \mbox{} - F_3G_3 \frac{M_2 m^2}{M_1^4} ( EE_\nu^0 + l^2 + {\mathbf p}_2 \cdot {\mathbf l}), \label{eq:q1}
\end{eqnarray}
\begin{eqnarray}
Q_2^\prime & = & (F_1^2+G_1^2) \frac{M_2}{M_1} (M_1-E_2) - (F_1^2-G_1^2) \frac{E_2 M_1-M_2^2}{M_1} + 2F_1G_1 \frac{M_2}{M_1} (E_\nu^0-E) \nonumber \\
&  & \mbox{} + (F_1G_2+F_2G_1) \frac{E_2}{M_1} (E_\nu^0-E) - (F_1G_2-F_2G_1) \frac{M_2}{M_1} (E_\nu^0-E) \nonumber \\
&  & \mbox{} - (F_1G_3+F_3G_1) \frac{E_2 m^2}{M_1^2} + (F_1G_3-F_3G_1) \frac{M_2 m^2}{M_1^2} - (F_1F_2+G_1G_2) \frac{p_2^2}{M_1}, \label{eq:q2}
\end{eqnarray}
$D_3=2(f_1^\prime g_1^\prime-{g_1^\prime}^2)$, and $D_4=2(f_1^\prime g_1^\prime+{g_1^\prime}^2)$. In Eqs.~(\ref{eq:aprime})-(\ref{eq:q2}) and hereafter $p_2$ and $l$ will denote the magnitudes of the corresponding 3-momenta. To avoid making the notation more cumbersome, we did not put primes on the form factors on the r.h.s.\ of Eqs.~(\ref{eq:q1}) and (\ref{eq:q2}). However, it must be kept in mind that it is the primed form factors $f_1^\prime$ and $g_1^\prime$, where all the model-dependence has been absorbed, that appear in these equations. We have limited ourselves to put primes on $Q_1^\prime$ and $Q_2^\prime$, as a reminder of this fact.

To the virtual RC of Eq.~(\ref{eq:dVp}) one must add the bremsstrahlung RC. It arises from the radiative decay
\begin{equation}
A \to B + \ell + \nu_\ell + \gamma,
\end{equation}
where $\gamma$ is a real photon with 4-momentum $k=(\omega,{\mathbf k})$ and which, in order to regulate the infrared divergence, is emitted with mass $\lambda$ and with an additional longitudinal degree of freedom. The summation over its polarization is performed according to Ref.~\cite{gins} and its model-independent contribution is controlled with the Low theorem \cite{low}. The integrations over $k$ are performed covariantly following Ref.~\cite{gins}.

Introducing the projector $\Sigma(s_2)$ in the radiative decay transition amplitude, following the usual steps of squaring it and summing over all polarizations including the summation over the $\gamma$ polarization \cite{gins}, and extracting the ${\hat {\mathbf s}_2}^B \cdot {\hat {\mathbf l}}$ correlation as explained in the virtual part, one obtains for the differential decay rate the result
\begin{equation}
d\Gamma_B = d\Gamma_B^\prime - d\Gamma_B^{(s)},
\end{equation}
where $d\Gamma_B^\prime$ is independent of ${\hat {\mathbf s}_2}^B$ and can be identified with one-half the unpolarized decay rate of Eq.~(27) of Ref.~\cite{mar01}. The $B$ spin-dependent part is given by
\begin{equation}
d\Gamma_B^{(s)} = \frac{\alpha}{\pi}d\Omega \left( I_0B^\prime + D_3 \rho_3 + D_4 \rho_4 \right) {\hat {\mathbf s}_2}^B \cdot {\hat {\mathbf l}}.
\end{equation}
$I_0(E,E_2)$ is the infrared-divergent integral given in Eq.~(26) of Ref.~\cite{mar01}, $B^\prime$ is identified with Eq.~(\ref{eq:bprime}) of this work. The contributions which are different with respect to the corresponding ones of Ref.~\cite{mar01} are
\begin{eqnarray}
\rho_3 & = & \frac{p_2l}{4\pi} \int_{-1}^{1} dx \int_{-1}^{y_0}dy \int_0^{2\pi} \frac{d\phi_k}{D(1-\beta x)}\left\{(p_2 y+l+\omega x) \left[ \frac{\beta^2 (1-x^2)}{1-\beta x} + \frac{\omega}{E} \right] \right. \nonumber \\
&  & \mbox{} \left. + \frac{\beta^2(1-x^2)}{1-\beta x} \left[ xE-\frac{D}{\beta} \right] \right\}, \label{eq:rho3}
\end{eqnarray}
and
\begin{eqnarray}
\rho_4 & = & \frac{p_2l}{4\pi} \int_{-1}^{1} dx \int_{-1}^{y_0}dy \int_0^{2\pi} \frac{d\phi_k}{D(1-\beta x)} \left\{ \left[ \frac{\beta^2(1-x^2)}{1-\beta x} \right] l \right. \nonumber \\
&  & \mbox{} \left. + E_\nu \left[ -\beta + \left( \frac{1-\beta^2}{1-\beta x} - 1 - \frac{\omega}{E} \right) x\right] \right\}, \label{eq:rho4}
\end{eqnarray}
where
\begin{equation}
D = E_\nu^0 + ({\mathbf p}_2 + {\mathbf l}) \cdot {\hat {\mathbf k}},
\end{equation}
and
\begin{equation}
\omega =\frac{p_2l (y_0-y)}{D}.
\end{equation}

The neutrino energy is $E_\nu=E_\nu^0-\omega$, where $E_\nu^0=M_1-E_2-E$, and $y_0=({E_\nu^0}^2-p_2^2-l^2)/2p_2l$. In Eqs.~(\ref{eq:rho3}) and (\ref{eq:rho4}) ${\mathbf k}$ is refered to a coordinate axis system where ${\hat {\mathbf l}}$
points along the $z$-direction and ${\hat {\mathbf p}_2}$ lies on the $(x,z)$ plane. The integration over ${\mathbf k}$ is performed with the variables $y={\hat {\mathbf p}_2} \cdot {\hat {\mathbf l}}$, $x={\hat {\mathbf k}} \cdot {\hat {\mathbf l}}$, and the azimuthal angle $\phi_k$. They are ready to be performed numerically.

In contrast with Ref.~\cite{neri08} where the integrals contained in the parts corresponding to Eqs.~(\ref{eq:rho3}) and (\ref{eq:rho4}) --namely, Eq.~(35) of this reference-- were most of them new and required a substantial effort to be performed analytically, all the integrals contained in Eqs.~(\ref{eq:rho3}) and (\ref{eq:rho4}) have already been performed analytically in our previous work. It is here where the parallelism between the present analysis and the one for the ${\hat {\mathbf s}_1} \cdot {\hat {\mathbf l}}$ calculation \cite{mar01} becomes very useful. One can compare Eqs.~(\ref{eq:rho3}) and (\ref{eq:rho4}) with their counterparts Eqs.~(34) and (35) of this last reference and observe that the integrals with the common factor $D_4$ correspond to the integrals in our $\rho_3$, while those with the common factor $D_3$ correspond to our $\rho_4$. Looking at the analytical results of the integrals in those Eqs.~(34) and (35), given in Eqs.~(46) and (47) of this same reference, we can establish the following connection
\begin{equation}
\rho_3^{{\hat {\mathbf s}}_2^B \cdot {\hat {\mathbf l}}} = (\rho_2^l + \rho_4^l )^{{\hat {\mathbf s}_1} \cdot {\hat {\mathbf l}}},
\end{equation}
and
\begin{equation}
\rho_4^{{\hat {\mathbf s}_2}^B \cdot {\hat {\mathbf l}}} = (\rho_1^l + \rho_3^l)^{{\hat {\mathbf s}_1} \cdot {\hat {\mathbf l}}},
\end{equation}
where $\rho_1^l,\dots,\rho_4^l$ are given explicitly in Eqs.~(48) and (49) of such Ref.~\cite{mar01}. Here the upper labels
${\hat {\mathbf s}}_2^B \cdot {\hat {\mathbf l}}$ and ${\hat {\mathbf s}_1} \cdot {\hat {\mathbf l}}$ are introduced to stress this correspondence. Using these results, and after some rearrangement to get somewhat more compact expressions, the analytical forms of Eqs.~(\ref{eq:rho3}) and (\ref{eq:rho4}) are
\begin{eqnarray}
\rho_3 & = & \frac{p_2}{2} \left[ E^2(Y_2-Y_3) - 2\theta_0E + Z_1 + \frac12 m^2 [2(1-\beta^2) \theta_2 - 5\theta_3] + \frac12 (3E^2-2l^2) \theta_4 \right. \nonumber \\
&  & \mbox{} \left. - \frac{3}{2}El\theta_5 - (1-\beta^2) \frac{E}{2} \theta_6 + \frac{3}{2}E\theta_7 + \frac{\theta_9}{4} + l^2\theta_{10} - \frac{l}{2} \theta_{14} - \frac12(4E+E_\nu^0) \eta_0 + \frac{\zeta_{21}}{2E} \right],
\end{eqnarray}
\begin{equation}
\rho_4 = \frac{p_2}{2} \left[ l^2Y_2 - \frac12(2E-E_\nu^0)\eta_0 - \frac12(E+2E_\nu^0) \gamma_0 + \frac{Y_4E}{2} + \frac{l^2}{2}\theta_3 \right].
\end{equation}
The explicit forms of the functions $\theta_i$, $\gamma_0$, $\eta_0$, $\zeta_{ij}$, $Y_i$ and $Z_1$ need not be reproduced here. They are all found in Ref.~\cite{torres04}.

Collecting the virtual and bremsstrahlung RC our final result is
\begin{equation}
d\Gamma \left( A^- \to B^0 e^- \overline{\nu} \right) = d\Omega \left\{ \left[ A_0^\prime + \frac{\alpha}{\pi} \Theta_I \right] - \left[ A_0^{(s)} + \frac{\alpha}{\pi} \Theta_{II} \right] {\hat {\mathbf s}_2}^B \cdot {\hat {\mathbf l}} \right\}, \label{eq:dgT}
\end{equation}
where the explicit forms of $\Theta_I$ and $A_0^\prime$ coincide with $\Phi_1$ of Eq.~(54) and $A_0^\prime$ of Eq.~(B1) of Ref.~\cite{mar01}, respectively. Our new results are $A_0^{(s)}$ of Eq.~(\ref{eq:aprime}) and
\begin{equation}
\Theta_{II} = B^\prime \left[ \phi + I_0(E,E_2) \right] + B^{\prime \prime} \phi^\prime + D_3\rho_3 + D_4\rho_4, \label{eq:fhis}
\end{equation}
where all the entries are defined above. The final result is infrared convergent.

We should recall that the practical application in the Monte Carlo analysis may be to use the RC in the form
\begin{equation}
\Theta_i = a_if_1^2 + b_if_1g_1 + c_ig_1^2, \label{eq:fhi}
\end{equation}
and to calculate the numerical values of the coefficients $a_i$, $b_i$, $c_i$ throughout the Dalitz plot in the form of arrays. Such arrays would be fed into the Monte Carlo simulation as a matrix multiplication. This procedure should save a substantial computer effort. In Eq.~(\ref{eq:fhi}) the index is $i=I,II$.

The present results complement the ones of Ref.~\cite{neri08}. The RC to the Dalitz plot when the ${\hat {\mathbf s}_2}^B \cdot {\hat {\mathbf l}}$ angular correlation is observed have been obtained both in the rest frame of the emitted baryon $B$ and in the rest frame of the decaying baryon $A$. They cover the three-body region of the Dalitz plot, they can be used for all charge assignments of $A$ and $B$ \cite{rfm02}, and $l$ may be $e^\pm$, $\mu^\pm$, or $\tau^\pm$. They are presented in a form which is not compromised to fixing the values of the form factors at prescribed values. They provide a good approximation to RC of medium- (several tens of thousands) and low- (several thousands) statistics experiments in light- and heavy-quark baryon semileptonic decays, respectively. As a final remark, let us stress that even if we use the same notation of our previous work, the expressions here apply only to the present case and there should arise no confusion.

The authors acknowledge financial support from CONACYT (M\'exico). J.J.T.\ and A.M.\ are grateful for partial support of COFAA-IPN (M\'exico). R.F.-M.\ also acknowledges financial support from FAI-UASLP (M\'exico).


\begin{thebibliography}{99}

\bibitem{neri08}
M.~Neri, J.~J.\ Torres, R.\ Flores-Mendieta, A.~Martinez, and A.~Garcia,
Phys.\ Rev.\  D \textbf{78}, 054018 (2008).

\bibitem{mar01}
A.~Martinez, J.~J.~Torres, R.~Flores-Mendieta and A.~Garcia,
Phys.\ Rev.\  D \textbf{63}, 014025 (2000).

\bibitem{gins71} E.\ S.\ Ginsberg, Phys.\ Rev.\ D \textbf{4}, 2893, (1971).

\bibitem{gins}
E.~S.~Ginsberg,
Phys.\ Rev.\ \textbf{162}, 1570 (1967)
[Erratum-ibid.\  \textbf{187}, 2280 (1969)].

\bibitem{low}
F.~E.~Low,
Phys.\ Rev.\  \textbf{110}, 974 (1958). H.\ Chew, Phys.\ Rev.\ \textbf{123}, 377 (1961).

\bibitem{torres04}
J.~J.~Torres, R.~Flores-Mendieta, M.~Neri, A.~Martinez and A.~Garcia,
Phys.\ Rev.\  D \textbf{70}, 093012 (2004)
[Erratum-ibid.\  D \textbf{75}, 019903 (2007)] and references therein.

\bibitem{rfm02}
R.~Flores-Mendieta, A.~Garcia, A.~Martinez and J.~J.~Torres,
Phys.\ Rev.\  D \textbf{65}, 074002 (2002).

\end{thebibliography}
\end{document}